\documentclass[article]{IEEEtran}
\usepackage[utf8]{inputenc}
\usepackage{url}
\usepackage{enumitem}
\usepackage{multirow} 
\usepackage{xcolor} 
\usepackage{graphicx}
\graphicspath{{figs/}}
\usepackage{hyperref}

\title{EPIC: An Energy-Efficient, High-Performance GPGPU Computing Research Infrastructure}

\author{Magnus Sj\"alander, Magnus Jahre, Gunnar Tufte, and Nico Reissmann \\
Norwegian University of Science and Technology (NTNU)\\
firstname.lastname@ntnu.no}

\begin{document}
\maketitle

\begin{abstract}
The pursuit of many research questions requires massive computational resources. State-of-the-art research in physical processes using simulations, the training of neural networks for deep learning, or the analysis of big data are all dependent on the availability of sufficient and performant computational resources. For such research, access to a high-performance computing infrastructure is indispensable. %

Many scientific workloads from such research domains are inherently parallel and can benefit from the data-parallel architecture of general purpose graphics processing units (GPGPUs). %
However, GPGPU resources are scarce at Norway's national infrastructure. %

EPIC is a GPGPU enabled computing research infrastructure at NTNU. It enables NTNU's researchers to perform experiments that otherwise would be impossible, as time-to-solution would simply take too long.
\end{abstract}

\section{Introduction}

The end of Dennard's scaling left computing systems across all domains increasingly power constrained. %
Specialized hardware in the form of accelerators emerged as alternatives to perform computations more energy-efficient. %
Specifically, general-purpose, graphic-processing units (GPGPUs) became increasingly popular as a means to accelerate programs in high-performance computing (HPC) and artificial intelligence. %

GPGPUs devote more compute resources to accelerate data-parallel applications by sacrificing resources that improve sequential program performance, rendering them more energy-efficient for data-parallel application domains. %
Nowadays, GPGPUs are significantly employed in High-Performance Computing (HPC) systems to meet performance demands while maintaining power constraints. %
For example, nine of the ten most powerful supercomputers in the world rely on GPGPUs for their computational power~\cite{top500}. %
Furthermore, eight of the top ten most energy-efficient supercomputers in the world rely on GPGPUs~\cite{green500}. %

The EPIC research infrastructure is a project between the Department of Computer Science and the IT Division at the Norwegian University of Science and Technology (NTNU) that aims at providing a GPGPU compute platform. %
EPIC is a part of the NTNU Idun computing cluster~\cite{Idun}, which provides a high-availability and professionally administrated compute platform for NTNU. %
Idun combines compute resources of individual shareholders to create a cluster for rapid testing and prototyping of HPC software. %
Currently, EPIC constitutes 48\% of the total number of nodes in the IDUN cluster and 100\% of the GPGPU resources. %

EPIC is with its 158 GPGPUs one of Norway's largest GPGPU enabled computational infrastructures. %
Norwegian national infrastructure has a very limited number of GPGPU resources, e.g., Saga~\cite{Saga} has only 32 NVIDIA Tesla~P100~\cite{Nvidia-P100} and Colossus~\cite{Colossus} has 32 much older NVIDIA Tesla~K20. %

\section{The Idun Cluster}

The Idun cluster is a Tier-2 \cite{prace} research cluster at NTNU meant as a stepping stone for the national infrastructure and serves as a platform for rapid testing and prototyping of HPC software, research into energy-efficient computing, and GPU-aided simulations and design-space exploration.

\begin{figure*}
  \centering
  \includegraphics[width=0.95\textwidth]{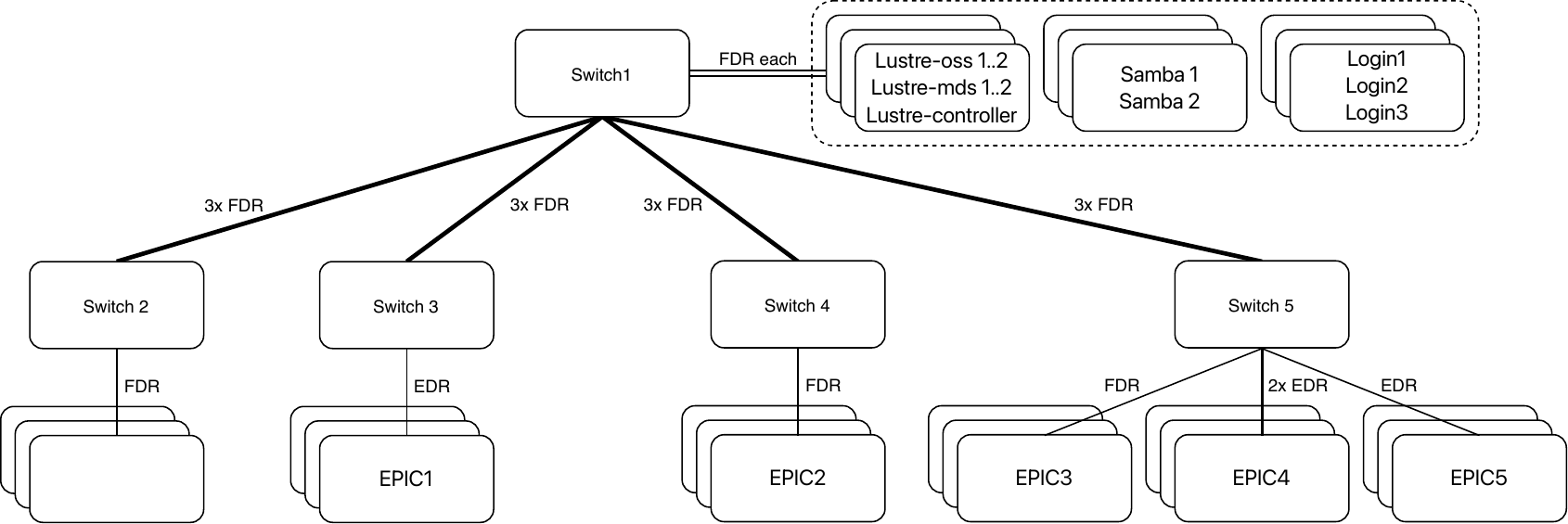}
  \caption{The topology of the Idun with the EPIC research infrastructure.}
  \label{fig:topology}
\end{figure*}

Currently, Idun consists of 73 nodes 
connected by two networks: one ethernet network and one high-throughput and low-latency InfiniBand (IB) network. The 1~Gb/s ethernet network serves as an administration and provisioning network, while the IB network is used for inter-node communication. The IB network is a mix of FDR (4x lanes each of 14~Gb/s) and EDR (4x lanes each of 25~Gb/s), as shown in \autoref{fig:topology}. Each node is connected with either FDR or EDR, resulting in 56~Gb/s or 100~Gb/s per node, respectively. The individual IB switches are connected in a tree structure with 3xFDR links between each switch, resulting in 168~Gb/s inter-switch connection speed.

Idun's storage is provided by two storage arrays and a Lustre parallel distributed file system~\cite{Lustre}. The storage arrays, one serves as Lustre metadata target (MDT) and one as Lustre object storage target (OST), are complemented with two Lustre metadata servers (MDS) and two object storage servers (OSS). The MDT and MDSs store the namespace data of the file system, such as filenames, directories, access permissions and file layouts, while the OST and OSSs store the file data. Together, the IB network and the Lustre file system, provide the means to efficiently transfer data to the compute resources, enabling an effortless
scaling of the cluster in terms of nodes and/or GPUs.


\section{The EPIC Research Infrastructure}

\begin{table*}
    \centering
    \caption{EPIC configuration}
    \label{tab:epic_configurations}
    \begin{tabular}{l|c|c|c|c|c|c|c|c}
         Name &  \#Nodes & Machine Model & \#CPUs & Processor model & \#Cores & Memory & \#Accel. & Accelerator model \\


         \hline
         EPIC1 & 
         8 & 
         Dell PE730 & 
         2 & 
         Intel Xeon E5-2695 v4~\cite{Intel-2695} & 
         36 & 
         128 GiB & 
         2 & 
         NVIDIA Tesla P100 16 GiB~\cite{Nvidia-P100} \\
         \hline
         EPIC2 & 
         19 & 
         Dell PE730 & 
         2 & 
         Intel Xeon E5-2650 v4~\cite{Intel-2650} & 
         24 & 
         128 GiB & 
         2 & 
         NVIDIA Tesla P100 16 GiB~\cite{Nvidia-P100} \\
         \hline
         EPIC3 & 
         5 & 
         Dell PE740 & 
         2 & 
         Intel Xeon Gold 6132~\cite{Intel-Gold-6132} & 
         28 & 
         768 GiB & 
         2 & 
         NVIDIA Tesla V100 16 GiB~\cite{Nvidia-V100} \\
         \hline
         \multirow{2}{*}{EPIC4} & 
         2 & 
         Dell DSS8440 & 
         2 & 
         Intel Xeon Gold 6148~\cite{Intel-Gold-6148} & 
         20 & 
         768 GiB & 
         8 & 
         NVIDIA Tesla V100 32 GiB~\cite{Nvidia-V100} \\
         \cline{2-9}
         
           & 
         1 & 
         Dell DSS8440 & 
         2 & 
         Intel Xeon Gold 6148~\cite{Intel-Gold-6148} & 
         20 & 
         768 GiB & 
         10 & 
         NVIDIA Tesla V100 32 GiB~\cite{Nvidia-V100} \\
         
        \hline
         \multirow{3}{*}{EPIC5} & 
         4 & 
         Dell DSS8440  & 
         2 & 
         Intel Xeon Gold 6148R~\cite{Intel-Gold-6148}  & 
         24 & 
         1.5 TiB & 
         10 & 
         NVIDIA A100-40~\cite{Nvidia-A100} \\ 
         \cline{2-9}
         
           &
         7 & 
         Dell XE8545  & 
         2 & 
         AMD EPYC 7543~\cite{AMD-EPYC-7543}  & 
         32 & 
         2 TiB & 
         4 & 
         NVIDIA A100-80~\cite{Nvidia-A100} \\ 
         \cline{2-9}
         
           & 
         2 & 
         Dell R740 & 
         2 & 
         Intel Xeon Gold 6132~\cite{Intel-Gold-6132} & 
         24 & 
         754 GiB & 
         4 & 
         Xilinx Alveo U250~\cite{Xilinx-U250} \\
    \end{tabular}
\end{table*}

The EPIC research infrastructure consists of five distinct investments (see \autoref{tab:epic_configurations}), each with a distinct purpose:

The original \textbf{EPIC1} consists of eight nodes with two NVIDIA P100 GPUs and focused on energy-efficient computing research such as energy efficient resource management for latency-critical cloud services~\cite{nishtala+:HPCA2020twig}.

\textbf{EPIC2} consists of 19 GPGPU nodes, each equipped with two NVIDIA P100 GPUs. %
These nodes complement EPIC1 and provide raw computational GPU power. %
These nodes are used for research in 3D object identification~\cite{bart:CG2020radial}, physical
simulations (e.g., nanomagnet ensemble dynamics modeled in MuMAX~\cite{vansteenkiste+:AIP2014mumax}
and flatspin~\cite{jensen2022flatspin}), and deep learning. %

\textbf{EPIC3} consists of five big-memory nodes, each equipped with two NVIDIA V100 GPUs. These nodes are meant for AI research that requires massive training sets, and therefore need more main memory.

\textbf{EPIC4} is an extension of EPIC2 providing another 26 GPUs for raw computational power. It consists of one node with ten V100 32 GiB GPUs and two nodes with eight V100 32 GiB GPUs. In addition, the big-memory GPUs (32 GiB instead if 16 GiB) enable larger working set sizes beneficial for 3D object identification and large AI models.



\textbf{EPIC5} is a further extension that adds 68 GPUs distributed across 11 nodes as well as four Xilinx Alveo U250 field programmable gate array (FPGA) accelerator cards~\cite{Xilinx-U250} distributed across two nodes. The primary use of the FPGA accelerators are to support computer architecture research using, e.g.,  FireSim~\cite{karandikar_firesim:_2018}.

Even though the the purpose and configuration of EPIC1-5 differ, all 158 GPGPUs can be accessed as one distributed resource for massive GPGPU performance.

\section{Research Outcome}

The EPIC cluster has been an indispensable resource for a wide range or research, e.g., efficient resource management, nanomagnetic modeling, 3D object identification, etc. Below is a non-exhaustive list of published articles that relied on EPIC 
to produce their results:
\begin{itemize}
    \item Energy-efficient resource management for latency-critical cloud services~\cite{nishtala+:HPCA2020twig}.
    \item Emergent computation on magnetic ensembles~\cite{jensen+:2018computation, lykkebo+:TEMC2019,
        jensen-tufte:ALIFE2020reservoir, jensen2022flatspin,
        penty2021representation, penty:ALIF2023, penty2023evolving, jensen:NATIURCOM2024}.
    \item Bit-serial matrix multiplication
      acceleration~\cite{Umuroglu+:TRETS2019bismo, metz+:TECS2023bisdu}.
    \item Intermediate representation (IR) for optimizing compilers~\cite{reissmann+:TECS2020rvsdg}.
    \item Nano-scale structures of aluminum  alloys~\cite{christiansen+:2019nano, christiansen+:2020detailed, christiansen+:microscopy2020}.
    \item Management of Internet of things (IoT) devices~\cite{Murad:2019iv}.
    \item Numerical modeling of renewable energy production and storage~\cite{vachaparambil2019comparison, vachaparambil2020numerical, vachaparambil2019spurious, vachaparambil2020sharp, jalili2018temperature, jalili2018new}.
    \item Bankruptcy prediction using machine learning~\cite{naess:2017konkurs, wahlstrom2016thesis}.
    \item Interest rate and treasury securities modeling~\cite{wahlstrom2020blissmodels, wahlstrom2021comparative}.
    \item Isogeometric analysis of acoustic scattering~\cite{venaas2019acoustic, venas2018isogeometric, venas2020isogeometric}.
    \item Framework for wind field predictions~\cite{tran_gans}.
    \item 3D object identification~\cite{bart:CG2020radial, van2021partial}.
    \item Computational fluid dynamics~\cite{toft:NIKT2020,  arosemena+:fluidmechnics2021, kozul2020aerodynamically, arosemena+:mechanics20221, arosemena+:2021effects}.
    \item Shear viscosity analysis~\cite{pousaneh+:2020kinetic, pousaneh+:2020shear}
    \item Molecular dynamics simulations~\cite{glende+:2020vanishing}
    \item Modeling of convection flows~\cite{Larkermani_Laurent:SINTEF2020}.
    \item Genetic association studies~\cite{halle+:2020computationally, aase+:2021biorxiv-genomic}.
    \item Behavior detection in echograms~\cite{maaloy2020echobert}.
    \item Sub-surface modifications~\cite{richter2020sub, richter2020investigation}.
    \item Autoignition-stabilized flames~\cite{gopalakrishnan2021}.
    \item Speculative side-channel
      mitigations~\cite{skalis+:SEED-iser, skalis+:SEED-dom,
        kvalsvik+:ISCA2023doppelganger, aimoniotis+:MICRO2023recon}.
    \item Modeling polymeric nanofibres~\cite{bering+:2021computational, bering+:2020entropy, bering:+2020legendre}
    \item Emission modeling~\cite{kramel+:EST2021}
    \item Modeling systemic circulation~\cite{bjordalsbakke:BIO2021parameter}
    \item Text processing using deep learning~\cite{turkerud2021image}
    \item Digital twins~\cite{sundby2021geometric}
    \item Neocortex encoding structure~\cite{mimica2022behavioral}
    \item Analytical models~\cite{torring2021autotuning}
    \item Population activity in grid cells~\cite{gardner2022toroidal}
    \item Cardinality constraints~\cite{hummel2021guaranteeing}
    \item Neuroscience~\cite{mimica2023behavioral}
    \item Computer graphics~\cite{sundt2023marf}
    \item Hyperspectral imaging~\cite{faltynkova2023developing}
    \item Performance profiling~\cite{gottschall+:MICRO2021tip,
        gottschall+:ISCA2023tea, gottschall:IISWC2023}
    \item Computational linguistics~\cite{kobzeva2023neural}
    \item Maintenance optimization~\cite{pedersen2023maintenance, pedersen2022optimizing}
    \item Uncertainty analysis~\cite{wang2023towards}
    \item Superconducting machines~\cite{hartmann2023static}
    \item Ship inspection~\cite{waszak2022semantic}
    \item Object anonymization~\cite{hukkelaas2023deepprivacy2}
    \item Spatial design~\cite{anyosa2023adaptive}
    \item Nonimaging optics~\cite{johnsen2022search}
    \item Numerical modeling of heat transfer~\cite{espelund2022numerical}
    \item Macromolecular crowding~\cite{blanco2021influence}
    \item Yield Curve Modeling~\cite{wahlstrom2021comparative}
    \item Stochastic search~\cite{mengshoel2021stochastic}
    \item Computer Security~\cite{sakalis+:TACO2022}
    \item And on many more topics~\cite{aspheim2024bayesian,
        aria2024full, haaversen2024qt, mellerud2024influence,
        naess2024bridging, klop2024electro, zhang2023comparison,
        dannert2023dna, faltynkova2024use, benestad2023efficient,
        tarekegn2023underwater, ullah2023multi, zhang2023computation,
        kalashnikov2023thermodynamics, mellerud2023design,
        moctezuma2022two, moctezuma2020eeg, cheng2023classification,
        hermansen2024uncovering, soler2022automated,
        moctezuma2020towards, saiti2022multimodal, borgelt2022native,
        sandal2022effects, fredriksen2022teacher, khalitov2022sparse,
        mendoza2022optimal, ringstad2022swirl, wiik2022unimolecular,
        arosemena2022characterization, cheng2024self, knuth2022semi,
        van2020indexing, cardaillac2022communication,
        lundregan2022dna, kobzeva2022lstms}
\end{itemize}

\subsection{PhD Theses}

The cluster has been used to produce results for at least the
following PhD theses:

\begin{enumerate}
\item Emil Christiansen, ``Nanoscale characterisation of deformed
  aluminium alloys'', 2019~\cite{christiansen:PhD2019nanoscale}.
\item Pablo Miguel Blanco, ``Coupling of binding and conformational
  equilibria in weak polyelectrolytes. Dynamics and charge regulation
  of biopolymers in crowded media.'', 2020~\cite{blanco:PhD2020coupling}.
\item Ranik Raaen Wahlstr{\o}m, ``Financial data science for exploring
  and explaining the ever-increasing amount of data'',
  2021~\cite{wahlstrom:PhD2021}.
\item Luis Alfredo Moctezuma ``Towards Universal EEG systems with
  minimum channel count based on Machine Learning and Computational
  Intelligence'', 2021~\cite{moctezuma:PhD2021}.
\item Eivind Bering, ``Stretching, breaking, and dissolution of
  polymeric nanofibres by computer experiments'',
  2021~\cite{bering:PhD2021}.
\item Jan Inge Hammer Meling, ``Hydrogen assisted crack growth in
  iron: a simulations approach'', 2021~\cite{meling:PhD2021}.
\item Christos Sakalis, ``Rethinking Speculative Execution from a
  Security Perspective'', Uppsala University,
  2021~\cite{sakalis:PhD_UU2021}.
\item Johannes H{\o}ydahl Jensen, ``Reservoir computing in-materio:
  Emergence and control in unstructured and structured materials'',
  2021~\cite{jensen:PhD2021}.
\item Ronja EM Wedeg{\"a}rtner, ``Highways up the mountains:
  Trails as facilitators for redistribution of plant species in
  mountain areas'', 2022~\cite{wedegartner2022highways}.
 \item Bj{\"o}rn Gottschall, ``Time-Proportional Performance Analysis
   for Out-of-Order Processors'', 2024~\cite{gottschall:PhD2024}.
 \item Anders Str{\o}mberg, ``Design and control of artificial spin
   ice'', 2024~\cite{stromberg:PhD2024}.
\end {enumerate}

\subsection{MSc Thesis Projects}

The cluster is also used as an educational resources where students
can run their simulations and produce results for their thesis
projects. %

\begin{enumerate}
  %
  %
\item Andr{\'e} H{\aa}land and Bj{\o}rnar Birkeland, ``Exploring data
  assignment schemes when training deep neural networks using data
  parallelism'', 2020~\cite{haaland2020exploring}.
\item J{\o}rgen Boganes, ``Accelerating Object Detection for
  Agricultural Robotics, 2020~\cite{boganes2020accelerating}.
\item Daniel {\O}rnes Halvorsen, ``Studies of turbulent diffusion
  through direct numerical simulation'',
  2020~\cite{halvorsen2020studies}.
\item Bj{\o}rn Magnus Valberg Iversen, ``Combining Hyperband and
  Gaussian Process-based Bayesian Optimization'',
  2020~\cite{iversen2020combining}.
\item Runar Ask Johannessen, ``Aggregation of Speaker Embeddings for
  Speaker Diarization'', 2020~\cite{johannessen2020aggregation}.
\item Siv-Marie McDougall, ``Fluorohectorite as a CO2 adsorbent: a DFT
  and DFTB study'', 2020~\cite{mcdougall2020fluorohectorite}.
  %
  %
\item M. Tarlton, Y. Roudi, and N. Bulso, ``Novel Model Selection
  Criterion for Inference of Ising Models'',
  2021~\cite{tarlton+:MSc2021}.
\item Richard Bachmann, ``Performance Modeling of Finite Difference
  Shallow Water Equation Solvers with Variable Domain Geometry'',
  2021~\cite{bachmann:MSc2021}.
  \item Frikk Hald Andersen and Eirik Dahlen, ``Sesame Street Pays
    Attention to Pro-Eating Disorder'',
    2021~\cite{andersen;Dahlen:MSc2021}.
  \item Martin Rebne Farstad, ``Understanding the Key Performance
    Trends of Optimized Iterative Stencil Loop Kernels on High-End
    GPUs'', 2021~\cite{farstad:MSc2021}.
\item Einar Aasli, ``Numerical Simulation of Fluid-Structure
  Interaction'', 2021~\cite{aasli:MSc2021}.
\item Klara Schl{\"u}ter and Jon Riege, ``Stochastic Multiplicative
  Updates for Symmetric Nonnegative Matrix Factorization'',
  2021~\cite{schluter_riege:MSc2021}.
\item Karoline Bonnerud, ``Write Like Me: Personalized Natural Language
  Generation Using Transformers'', 2021~\cite{bonnerud2021write}.
\item Michael Tarlton, ``Novel Model Selection Criterion for Inference
  of Ising Models'', 2021~\cite{tarlton2021novel}.
\item Anja Rosvold From and Ingvild Unander Netland, ``Fake News
  Detection by Weakly Supervised Learning: A Content-Based Approach'',
  2021~\cite{from2021fake}.
\item Didrik Salve Galteland, ``Exploring Self-supervised
  Learning-based Methods for Monocular Depth Estimation in an
  Autonomous Driving Setting'', 2021~\cite{galteland2021exploring}.
\item Vebj{\o}rn Malmin and Halvor{\O}deg{\aa}rd Teigen,
  ``Reinforcement Learning and Predictive Safety Filtering for
  Floating Offshore Wind Turbine Control'',
  2021~\cite{malmin2021reinforcement}.
\item Marthe Strand Haltbakk, ``Kinetic Monte Carlo simulation of the
  early precipitation stages in Al-Mg-Si alloys using Cluster
  Expansion methods for energy barrier modelling'',
  2021~\cite{haltbakk2021kinetic}.
\item Andreas Herl{\o}sund S{\o}gnen, ``Numerical analysis of
  finned-tubes and finned-tube bundles'',
  2021~\cite{sognen2021numerical}.
\item Aurora  Grefsrud, ``Efficiency of IllustrisTNG in modeling
  galaxy properties'', 2021~\cite{grefsrud2021efficiency}.
\item Joakim Olsen, ``Measuring Summary Quality using Weak
  Supervision'', 2021~\cite{olsen2021measuring}.
\item Varun Loomba and Jan Erik Olsen and Kristian Etienne Einarsrud,
 ``Modelling of Furnace Tapping with Uniform and Non-Uniform Porosity
 Distribution'', 2021~\cite{loomba2021modelling}.
\item Lars Andreas Hastad Lervik, ``Orientation and Projection Center
 Refinement for EBSD Indexing in Python'',
 2021~\cite{lervik2021orientation}.
\item Vemund Fredriksen and Svein Ole Matheson Sevle, ``Pulmonary
  Tumor Segmentation Utilizing Mixed-Supervision in a Teacher-Student
  Framework'', 2021~\cite{fredriksen2021pulmonary}.
\item Halvor Bakken Smed{\aa}s, ``ASSIST: Accuracy-driven Sampling
  Strategies for Improved Supervised Training'',
  2021~\cite{smedaas2021assist}.
\item Robin Christian Staff, ``What a Twist-Using Deep Neural Networks
 to Generate Plot Twists'', 2021~\cite{staff2021twist}.
\item H{\aa}kon S{\o}rensen B{\o}ckman, ``Locating sheep in the
 highlands with aerial footage and a lightweight algorithm system'',
 2021~\cite{sorensen2021locating}.
\item Alexander Michael Staff, ``An Empirical Study on Cross-data
 Transference of Adversarial Attacks on Object Detectors'',
 2021~\cite{staff2021empirical}.
\item Jon Steinar Folstad, ``Transformer Pre-Trained Language Models
 and Active Learning for Improved Blocking Performance in Entity
 Matching'', 2021~\cite{folstad2021transformer}.
\item Christian Ziegenhahn Jensen and Espen S{\o}rhaug, ``The
 Perfect Rap Lyrics-AI Generated Rap Lyrics That Are Better Than
 Lyrics from Existing Popular and Critically Acclaimed Rap
 Songs'', 2021~\cite{jensen2021perfect}.
\item Jostein Lillel{\o}kken and Martin Hermansen, ``Improving
 Performance of Autonomous Driving in Simulated Environments
 Using End-to-End Approaches'',
 2021~\cite{lillelokken2021improving}.
\item Michael Moen Allport and Jonas Sandberg, ``Q-PRM-A QoS Aware
 Resource Manager for Colocated Services'', 2021~\cite{allport2021q}.
\item Vilde Roland Arntzen, ``Detecting Norwegian Abusive Language in
 Social Media with Transformer-based Models'',
 2021~\cite{arntzen2021detecting}.
\item Magnus Midtb{\o} Kristiansen, ``Proving Theorems Using Deep
 Learning'', 2021~\cite{kristiansen2021proving}.
\item Emil Alvar Myhre, ``Bayesian optimal experimental design for
 studying synaptic plasticity'', 2021~\cite{myhre2021bayesian}.
  %
  %
\item Ingebrigt Nyg{\aa}rd and Sebastian Vitters{\o}, ``Improved
  Sheep Detection-Modifying YOLOv5 to accurately detect grazing
  sheep in UAV imagery'', 2022~\cite{nygaard2022improved}.
\item Fabian Vakhidi, ``Pose Estimation with Convolutional Neural
  Networks'', 2022~\cite{vakhidi2022pose}.
\item Andr{\'e}s Javier Est{\'e}vez Fern{\'a}ndez, ``Combining
  question answering models with transformer-based generative
  conversational agents'', 2022~\cite{estevez2022combining}
\item Jonas Strand Aasberg, ``Machine Learning using High Resolution
  Zivid Point Clouds on a High Performance Cluster'',
  2022~\cite{aasberg2022machine}.
\item Jenny Bleken Hellerud, ``{AI} and Emotions in Perfect
  Harmony-Recognising Emotions in Music using CQT Spectrograms and
  Multi-output Regression'', 2022~\cite{hellerud2022ai}.
\item Eivind Aksnes Rebnord, ``Generating Audio from Sample
  Librarie'', 2022~\cite{rebnord2022generating}.
\item Hallvard Stemshaug, ``Impact of Low Resolution IR Images in
  Drone Based Sheep Detection'', 2022~\cite{stemshaug2022impact}.
\item Christopher Michael Vibe, ``Practical Reservoir Computing \&
  Echo State Property Metrics'', 2022~\cite{vibe2022practical}.
\item Fredrik Alm{\aa}s, ``Bottom-detection in Doppler Velocity
  Logs using Recurrent Neural Networks on an embedded platform'',
  2022~\cite{almaas2022bottom}.
\item Angqi Zhao, ``Lithium cations mobility in a coarse-grained
  polymer embedded with Lennard Jones particles using
  non-equilibrium molecular dynamics'',
  2022~\cite{zhao2022lithium}.
\item Andr{\'e} Storhaug, ``Secure Smart Contract Code Synthesis
  with Transformer Models'', 2022~\cite{storhaug2022secure}.
  %
  %
\item Preben Gjelsvik, ``Optical Properties of Intermediate Band
  Ca6FeN5 and Related Materials: A Density Functional Theory Study'',
  2023~\cite{gjelsvik2023optical}.
\item Kristin Fr{\o}ystein, ``Scanning precession electron
  diffraction data analysis of in-situ precipitate evolution in an
  Al-Mg-Si-Cu alloy'', 2023~\cite{froystein2023scanning}.
\item Tobias Meyer Andersen, ``Performance Modeling of a Load-Balanced
  FDM Wave Equation Solver on Heterogeneous Clusters'',
  2023~\cite{andersen:MSc2023}.
\item Ole Joachim Arnesen Asen, ``Small languages and big models-Using
  ML to generate social media content for training purposes'',
  2023~\cite{aasen:MSc2023}.
\item Alexander Michael {\AA}s, ``A Norwegian Whisper Model for
  Automatic Speech Recognition'', 2023~\cite{aas:MSc2023}.
\item Karoline Lillevestre Langli, ``Sentiment Analysis of Customer
  Emails Using BERT'', 2023~\cite{langli:MSc2023}.
\item John Askeland Lauvdal, ``Investigating Speculative Side-Channel
  Protection'', 2023~\cite{lauvdal:MSc2023}.
\item Ulrik Bernhardt Danielsen, ``A step toward model selection in
  unsupervised clustering of animal behavior'',
  2023~\cite{danielsen:MSc2023}.
\item Velte Kristaver Widnes Harnes, ``Exploring Efficient
  Accelerator-Core Integration Strategies: A Case Study of BISMO in
  Chipyard'', 2023~\cite{harnes:MSc2023}.
\item Andreas R{\"o}nnestad, ``Evaluation of Safety-Oriented Metrics
  For Object Detectors'', 2023~\cite{ronnestad:MSc2023}.
  \item Karoline Seljevoll Herleiksplass, ``Enhancing Sleep-Wake
    Detection Using Deep Learning and Optimal Channel Selection from
    High-Density EEG'', 2023~\cite{herleiksplass:MSc2023}.
  \item Christopher Str{\o}m, ``Towards robust and flexible
    point-object multi-target tracking using transformer neural
    networks'', 2023~\cite{strom:MSc2023}.
  \item Marcus Stensby Young, ``Improving Memory Scheduling of an
    Out-of-Order Core'', 2023~\cite{young:MSc2023}.
  \item Clemens Martin M{\"u'}ller, ``Physics informed neural
      networks in radial load flow calculations'',
      2023~\cite{muller:MSc2023}.
\end{enumerate}

\section{Conclusion}

EPIC is a multi-million investment by the Department of Computer Science in collaboration with the IT Division to provide GPGPU resources for NTNU's researchers. %
The large number of GPGPUs enable research studies to be performed at a scale that  otherwise would be impossible to conduct. %
Thus, EPIC's computational resources help NTNU's researchers to stay competitive and produce state-of-the-art results. %

\section*{Acknowledgement}

The original EPIC1 was financed by an NTNU Advanced Research Equipment grant (AVIT) and EPIC3 was financed by the NTNU-Telenor AI Lab (now called the Norwegian Open AI Lab). EPIC2 and EPIC4 were financed by the Department of Computer Science. The GPU-nodes in EPIC5 were jointly funded by an AVIT grant, the Department of Computer Science and the Department of Electric Power Engineering  while the FPGA-nodes were jointly funded by the Department of Computer Science and incentive funds allocated to Magnus Jahre as a result of being awarded the BAMPAM project by the Research Council of Norway (grant no.\ 286596). The network, storage, and maintenance of the cluster is provided by the HPC group of NTNU's IT Division. 

\bibliographystyle{IEEEtranS}
\bibliography{defs,refs}

\end{document}